\documentclass[a4paper,11pt]{article}
\pdfoutput=1 

\usepackage{jinstpub} 

\title{\boldmath  Neutrinoless double-beta  decay search with the LEGEND experiment}

\author[a]{R. Brugnera}

\affiliation[a]{Universit\'{a} degli Studi di Padova and Padova INFN, \\
   via Marzolo 8, Padova, Italy}

\emailAdd{brugnera@pd.infn.it}

\abstract{Neutrinoless double-beta decay is a nuclear decay, given as $(A,Z) \rightarrow (A, Z+2) +2e^{-}$, with deep consequences for the understanding of our universe.
A strong experimental program is underway to search for this transition with many proposed experiments using different technologies. In this article the LEGEND experiment, which uses $^{76}$Ge as the isotope of interest, will be described. We will discuss both the first stage, LEGEND-200,  which is now taking data at the Laboratori Nazionali del Gran Sasso of INFN in Italy, and the future stage,  LEGEND-1000. LEGEND-200 has analyzed a first sample of data (48.3 kg$\cdot$yr) collected from March 2023 to February 2024 with a background index not far away from its goal of 2$\times$10$^{-4}$ cnts/(keV$\cdot$kg$\cdot$yr). Combining the LEGEND-200 data with those of \textsc{Gerda}\ and \textsc{Majorana Demonstrator}\ one obtains a sensitivity  on the half-life of  0$\nu\beta\beta$ decay in  $^{76}$Ge of $T_{1/2} > $  2.8 $\times$ 10$^{26}$ yr at 90\% C.L. and a limit on  $T_{1/2}  > $  1.9 $\times$ 10$^{26}$ yr at 90\% C.L. 
}


\collaboration[c]{on behalf of the LEGEND collaboration}

\proceeding{LIDINE 2024: LIght Detection In Noble Elements\\
  Aug 26 - 28, 2024\\
 Instituto Principia - S\~{a}o Paulo - Brazil }

\begin{document}
\maketitle
\flushbottom

\section{Introduction}
\label{sec:intro}

\begin{figure}
\begin{center}
\includegraphics[width=7.5cm]{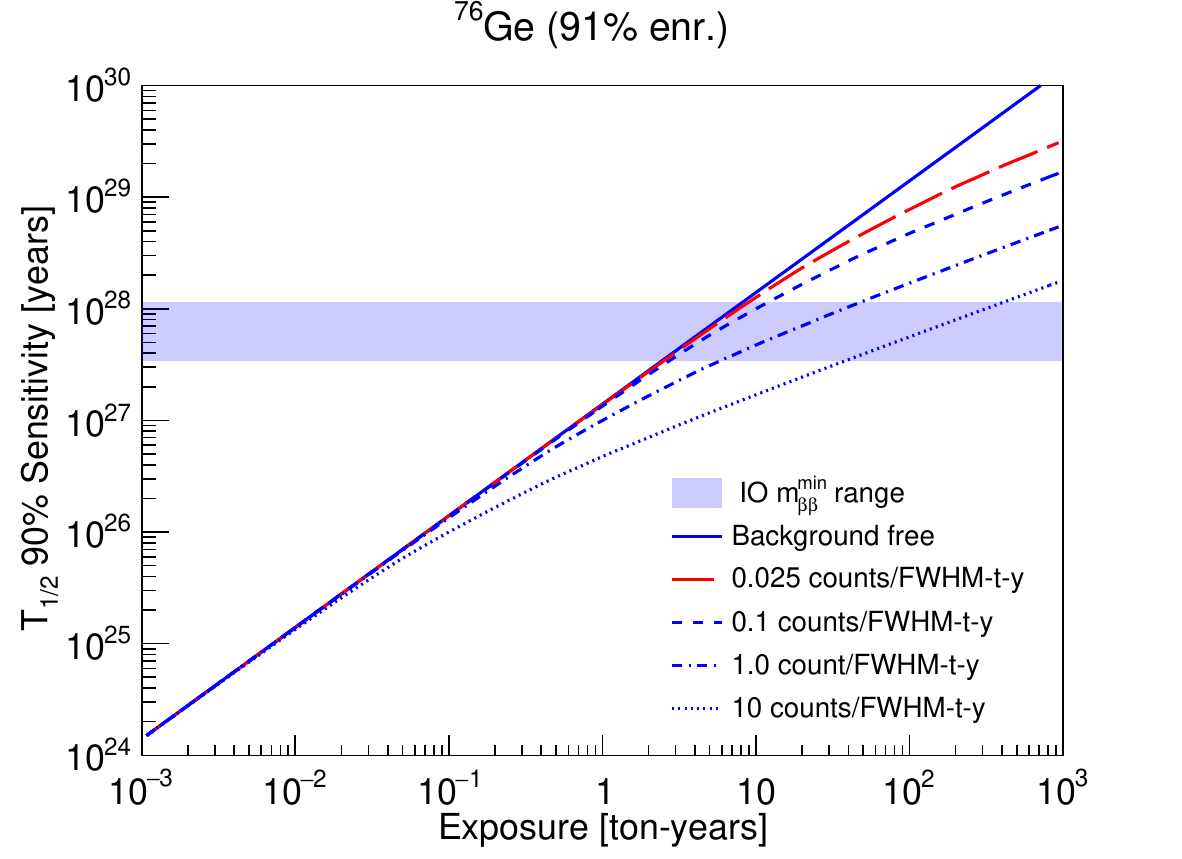}
\includegraphics[width=7.5cm]{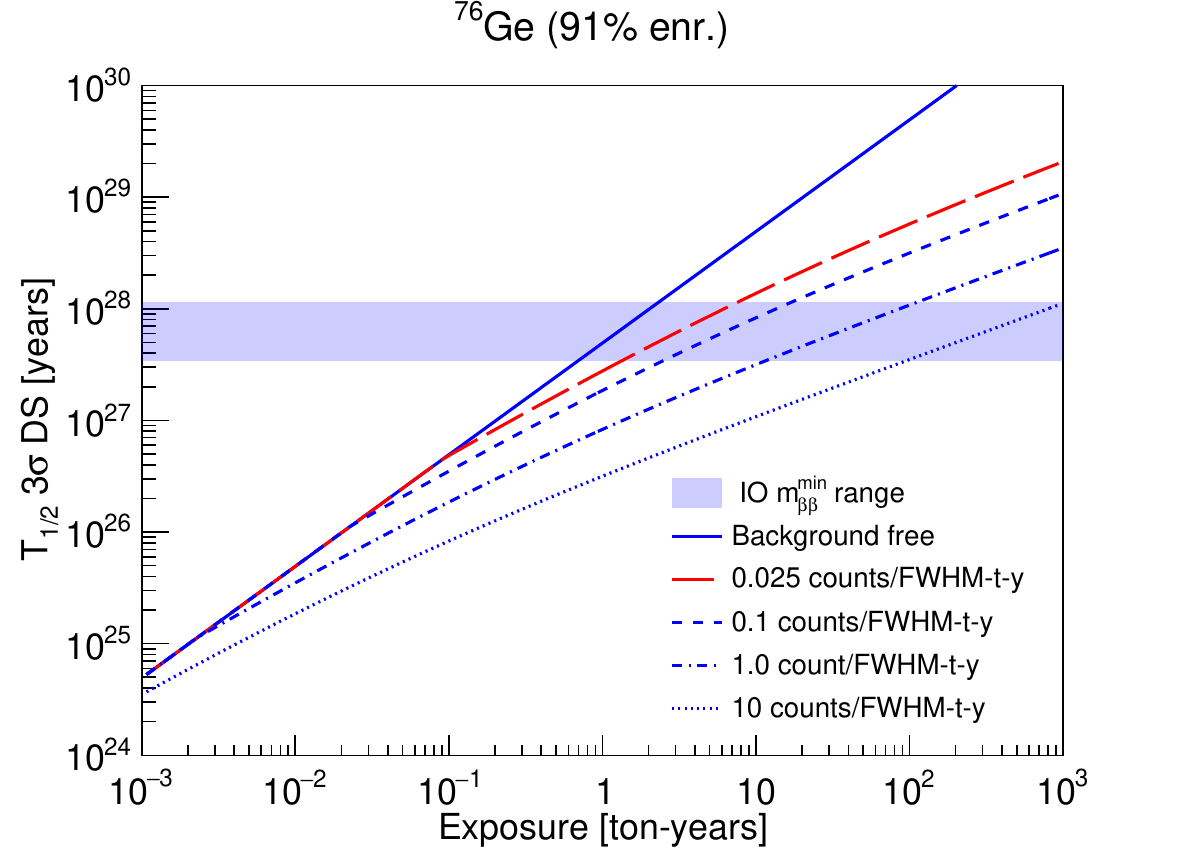}
\end{center}
\caption{Sensitivity for setting a limit (left) and discovery (right) as a function of the exposure and background. The background rates uses a 2.5 keV FWHM energy resolution around the $Q_{\beta\beta}$  of the reaction. The horizontal band corresponds to the range of half-lives, $T_{1/2}$, necessary to cover the lowest value of m$_{\beta\beta}$ permitted by the inverted order hierarchy. It's clear from these figures, that for a discovery to be made, background reduction is as important as exposure.}
\label{fig:sensitivity}
\end{figure}

Neutrinoless double-beta (0$\nu\beta\beta$) decay is a process that violates lepton number conservation by two units. Its observation would have other far-reaching consequences. For example, it would prove that neutrinos have a Majorana mass component. An effective Majorana neutrino mass (m$_{\beta\beta}$) can be connected to the decay half-life by using calculations for the nuclear matrix element, assuming the exchange of light Majorana neutrinos.
The isotope candidates for such nuclear transitions are even-even nuclei in which a single beta decay is energetically forbidden. 
A particularly promising isotope is $^{76}$Ge (with a total energy release of $Q_{\beta\beta}$\ = 2039.061$\pm$ 0.007 keV). Many experiments in the past have used this isotope and in recent years the two experiments \textsc{Gerda}\ \cite{bib:gerda} and \textsc{Majorana Demonstrator}\ \cite{bib:mjd} obtained competitive half-life limits of 1.8$\times$10$^{26}$ yr and 0.8$\times$10$^{26}$ yr at 90\% C.L., respectively. 
Building on the successes of and  the best technologies developed by \textsc{Gerda}\ and \textsc{Majorana Demonstrator}, the LEGEND \cite{bib:legend} collaboration aims to develop a phased  
0$\nu\beta\beta$  experimental program. LEGEND-200 is its first phase with the aim of reaching a sensitivity of about 10$^{27}$ yr 
in terms of both setting a 90\% C.L. limit and achieving a 50\% chance to make a 3 $\sigma$ discovery,
thanks to a projected background index of 
0.6 cnts/(FWHM$\cdot$t$\cdot$yr), or 2$\times$10$^{-4}$ cnts/(keV$\cdot$kg$\cdot$yr),  and an exposure of 1 t$\cdot$yr. The second phase, LEGEND-1000, aims for a sensitivity of beyond 10$^{28}$ yr by operating 1 tonne of enriched germanium detectors for an exposure of more than 10 t$\cdot$yr at a background index of about 0.025 cnts/(FWHM$\cdot$t$\cdot$yr), or 10$^{-5}$ cnts/(keV$\cdot$kg$\cdot$yr). Figure \ref{fig:sensitivity} shows the sensitivities for setting limits and for discovery potential for a $^{76}$Ge experiment  as a function of the exposure for different background indices. 

\section{The LEGEND-200 phase}
\label{sec:legend-200}

The LEGEND-200 experiment is located at the Laboratori Nazionali del Gran Sasso (LNGS) of INFN where a rock overburden of 3500 m water equivalent reduces the cosmic muons flux by 6 orders of magnitude. The heart of the experiment is presently made of 140 kg of germanium detectors of different types: Broad Energy Germanium (BEGe), Inverted-Coaxial Point-Contact (ICPC),  P-type Point Contact (PPC) and semicoaxial. The detectors are made from isotopically modified material with $^{76}$Ge enriched up to 92\% and are organized in 10 strings. The strings are lowered in a cryostat containing 64 m$^3$ of liquid argon 
(LAr) through a lock system inside a clean room. The cryostat is immersed in a water tank (containing 590 m$^3$ of purified water) equipped with photomultipliers to detect the residual cosmic muons reaching the experiment. The strings are surrounded by an inner and outer curtain of wavelength-shifting fibers connected to silicon photomultipliers for the detection of the scintillation light emitted by the LAr.

\begin{figure}
\begin{center}
\includegraphics[width=15.0cm]{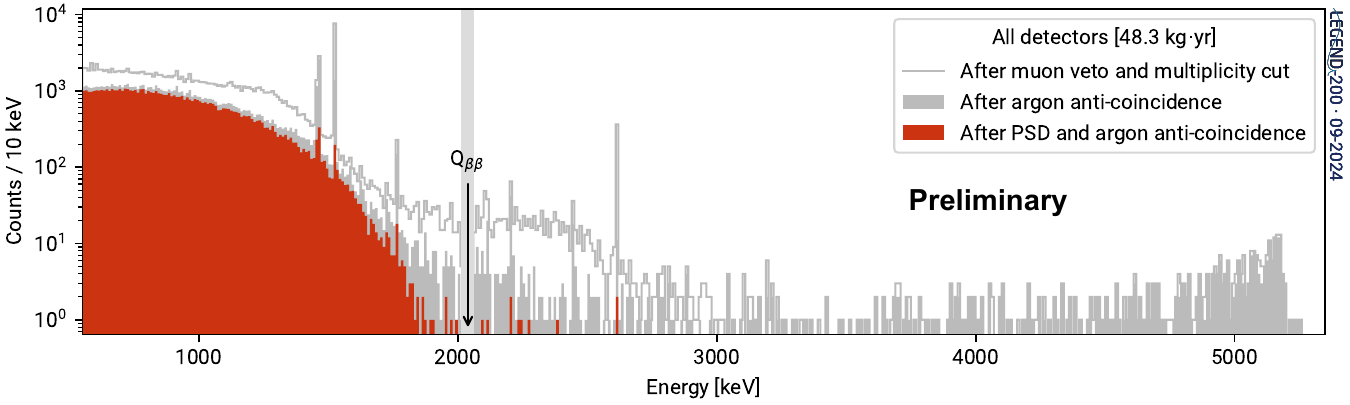}
\end{center}
\caption{Energy distribution of events between 565 and 5300 keV  prior to liquid argon veto and PSD cuts (total histogram), after LAr veto (light gray) and after all cuts (red). The gray vertical band indicates the blinded region of $\pm$ 25 keV around the $Q_{\beta\beta}$ value. The exposure is 48.3 kg$\cdot$yr.}
\label{fig:energy_spectrum}
\end{figure}

\begin{figure}
\begin{center}
\includegraphics[width=7.5cm]{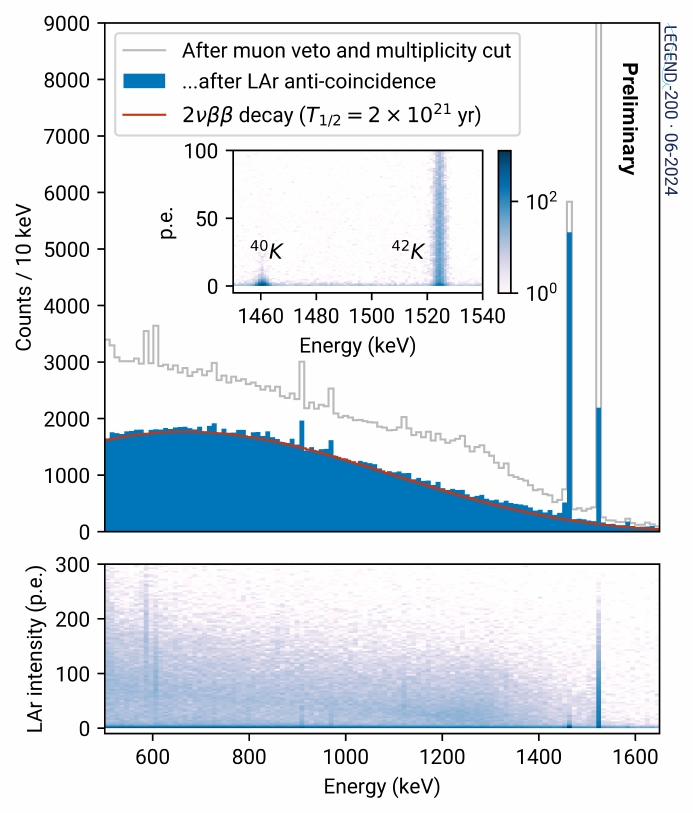}
\includegraphics[width=7.5cm]{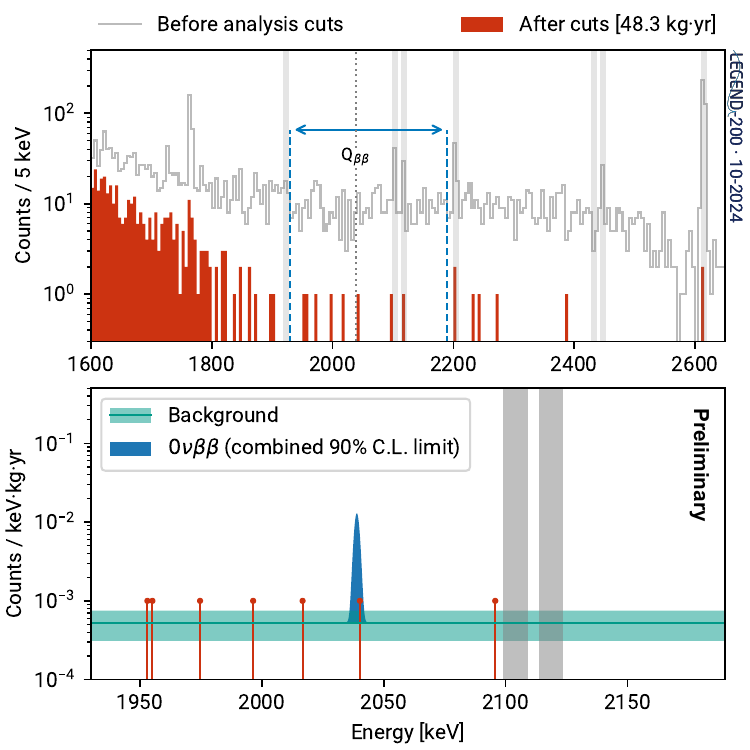}
\end{center}
\caption{Left upper plot: energy distribution of events between 565 keV and 1620 keV prior to LAr  anti-coincidence and PSD cuts (total histogram) and  after LAr anti-coincidence cut (full blue). The red curve represents the 2$\nu\beta\beta$ contribution normalized using the half-life measured by \textsc{Gerda}\ \cite{bib:gerda2nu}. The inset shows the photo-electrons detected by the LAr instrumentation in the energy region around the two potassium $\gamma$ lines. Left lower plot: The photo-electrons distribution as a function of the energy (from 565 keV to 1620 keV) registered in the Ge detectors. Right upper plot: Enlarged view of the energy distribution of events between 1.6 and 2.65 MeV prior to (total histogram)  and after all cuts (red). This energy interval includes the analysis window (edges marked by dashed lines) and the regions of expected $\gamma$ lines (marked by gray areas).
Right lower plot: result of the frequentist unbinned extended likelihood fit; the blue peak displays the expected 0$\nu\beta\beta$ decay signal for T$_{1/2}$ equal to the combined lower limit, 1.9 $\times$ 10$^{26}$ yr. Red vertical lines indicate the energies of the surviving events in the analysis window after all the analysis cuts.}
\label{fig:final_energy_spectrum}
\end{figure}

The first data taking period of LEGEND-200 began in March 2023 and ended in February 2024. 
 A total of 76.2 kg$\cdot$ yr of exposure  (silver data set) 
was collected and used for background and performance characterization. From this exposure, a subset of 48.3 kg$\cdot$ yr (golden data set) 
was selected for the following 0$\nu\beta\beta$ analysis. At present only BEGe, PPC and ICPC detectors are used;
 work is on-going to increase the golden data set. For the entire data taking the energy resolution of all types of Ge detectors remained stable  at ~0.1\% FWHM at the $Q_{\beta\beta}$\ value. 
The peculiar topology of signal-like events allows for most background events to be discarded.
Signal-like events release energy in a very narrow  region in a single germanium detector; for this reason they are called single-site events (SSE). In contrast, background events release energy either in more than one detector (eliminated by requiring energy release in a single detector, multiplicity cut), or in many different sites (multi-site events, MSE) of a single detector or on its surfaces (both these two last classes of events are eliminated using dedicated pulse shape discrimination (PSD) techniques), or they are accompanied by energy release in the LAr (detected by reading the LAr scintillation light).  
The total histogram in Figure \ref{fig:energy_spectrum} shows the energy spectrum after quality cuts, muon veto and multiplicity cut. The light gray and red histograms show the remaining events after the argon anti-coincidence cut is applied and finally after PSD and argon anti-coincidence cuts. These active rejection tools are particularly effective in eliminating background events around the $Q_{\beta\beta}$\ region. 
The top left plot of Figure \ref{fig:final_energy_spectrum} shows the energy spectrum from 565 keV to 1620 keV prior to the analysis cuts (total histogram) and then after the argon anti-coincidence cut (blue histogram). After this cut the spectrum is strongly dominated by 2$\nu\beta\beta$ events 
as shown by the red curve representing its predicted contribution using the 2$\nu\beta\beta$ half-life measured by \textsc{Gerda}\ \cite{bib:gerda2nu}.
The right top plot of Figure \ref{fig:final_energy_spectrum} shows a zoom of the spectrum around the  $Q_{\beta\beta}$\ value, the energy window from 1930 to 2190 keV (edges marked by dashed lines) is the analysis window where the 0$\nu\beta\beta$ statistical analysis is performed. 
Only 7 events remain in this region (right bottom plot of Figure \ref{fig:final_energy_spectrum}) of which one is at only 1.4 $\sigma$ from the $Q_{\beta\beta}$\ value, giving a background index of about 5$\times$10$^{-4}$ cnts/(keV$\cdot$kg$\cdot$yr).
A  frequentist fit combining the exposures of \textsc{Gerda}, \textsc{Majorana Demonstrator}\ and LEGEND-200 was done giving a limit of 1.9$\times$10$^{26}$ yr at 90\% CL and a median sensitivity greater than 2.8$\times$10$^{26}$ yr at 90\% CL.

Data taking was interrupted in February 2024 because a higher than expected level of background events was discovered in the analysis of the data around the $Q_{\beta\beta}$\ value. Most of 2024 was spent understanding the origin of these events through dedicated data taking and radio-assay campaigns of pieces of material placed near the Ge detectors. Data taking will be resumed in early 2025 with an apparatus with enhanced performance.

\section{The LEGEND-1000 phase}
\label{sec:legend-1000}
The next phase, LEGEND-1000 \cite{bib:legend-1000}, will consist of 1000 kg of ICPC detectors enriched to more than 90\% in $^{76}$Ge 
 operated in a liquid argon active shield at an underground laboratory. The baseline design assumes LNGS as host site.  Its goal is to fully explore the neutrino inverted order hierarchy down to  m$_{\beta\beta}$\ in the range  9-21 meV, in 10 years of live time, with a sensitivity beyond 10$^{28}$ years. 
Next year the experiment will start the path towards its approval process in the Critical Design-1 review organized by the DOE. 

\section{Conclusions}
\label{sec:conclusions}
We presented the LEGEND program that will search for 0$\nu\beta\beta$ decay with half-lives beyond 10$^{28}$ years in a staged approach.
LEGEND-200, its first stage, is now taking data with the lowest background level of any 0$\nu\beta\beta$ experiment. 
The future stage,  LEGEND-1000, is well under way,  with participation in the DOE Critical Design-1 review in the coming year.

\end{document}